# Vendor-agnostic 4D Phase Contrast MRI: a complete open-source pipeline for velocities, displacement, and strain analysis

**Marta B. Maggioni[1,2], Sabine M. Räuber[1,2],Katarina Puš[3,4], Bostjan Šimunič[3], Xeni Deligianni[1,5], Regina M. M. Schlaeger[6,7], Francesco Santini[1,2]**

*martabrigid.maggioni@unibas.ch, sabine.raeuber@unibas.ch, katarina.pus@zrs-kp.si, bostjan.simunic@zrs-kp.si, xeni.deligianni@unibas.ch, regina.schlaeger@usb.ch, francesco.santini@unibas.ch*

[1] Basel Muscle MRI (BAMM), Department of Biomedical Engineering, University of Basel, Basel, Switzerland.
[2] Radiology, Division of Radiological Physics, University Hospital of Basel, Basel, Switzerland.
[3]Institute for Kinesiology Research, Science and Research Centre Koper, Koper, Slovenia
[4]Department of Health Sciences, Alma Mater Europaea University, Maribor, Slovenia
[5]Paediatric Research Centre, Universitty Children's Hospital, University of Basel, Basel, Switzerland
[6] Department of Clinical Research, University of Basel, Basel, Switzerland
[7] Neurologic Clinic and Policlinic, University Hospital Basel, Basel, Switzerland

Corresponding author: **Marta B. Maggioni,**
email: martabrigid.maggioni@unibas.ch

Keywords: phase contrast, Muscle, Dynamic, Open-source, vendor-agnostic

## Abstract

Phase contrast MRI (PC MRI) enables quantitative assessment of tissue motion and strain. Although it is increasingly used, standardized, vendor-agnostic pipelines for accelerated acquisitions remain scarce. We present a fully open-source 4D flow PC-MRI pipeline integrating a compressed sensing-accelerated sequence implemented in PyPulseq, BART-based reconstruction, and strain analysis. Additionally, a gradient probing sequence was developed to ensure correct velocity sign assignment across scanner orientations and vendors. The pipeline was validated across two Siemens MRI systems (3T MAGNETOM Prisma and 3T Vida Fit) in two anatomical applications: forearm (Flexor Digitorum Superficialis, n=9) and thigh (Vastus Lateralis, n=10) during Neuromuscular Electrical Stimulation (NMES)-induced contractions. Compressed sensing reduced acquisition times from 35 and 80 minutes to 5 and 11 minutes for the arm and leg acquisitions, respectively. Muscle strain maps and sigmoid-fitted strain curves enabled extraction of peak strain, mean strain, and buildup rate. Strains in the Vastus Lateralis were approximately one order of magnitude higher than in the Flexor Digitorum Superficialis (median peak strain 0.49 vs. 0.063, mean strain 0.31 vs. 0.031). The pipeline demonstrates multi-platform compatibility and provides a reproducible, open framework for quantitative muscle imaging.

## Introduction

Blood and tissue motion can be accurately quantified with MRI using the phase contrast (PC-MRI) method, which has found broad application in cardio- and neurovascular, neurofluidic[1], and musculoskeletal (MSK) imaging. Three-dimensional, three-directional, time-resolved acquisitions (usually termed “4D-Flow”)[2] can measure complex motion patterns and can be used to evaluate tissue strain, an important parameter for the evaluation of cardiac and skeletal muscle tissues. However, the high data dimensionality requires long acquisition times unless advanced acceleration strategies are used, but standardized implementations of advanced acquisition, reconstruction, and processing pipelines are scarcely available, proprietary, and often different across scanner vendors.

PC-MRI relies on the addition of motion-sensitizing gradients to a conventional gradient echo acquisition, which allows velocity quantification along the axis of the gradient, and it is generally acquired in a gated or triggered fashion, relying on periodic flow or motion patterns. It was originally developed for cardiac applications but has since been applied to dynamic muscle imaging, where it has been successfully used to study muscle dynamics during both voluntary [3], [4] and externally induced contractions[5], [6] (through the use of electrical muscle stimulation). The latter method offers advantages as the contractions are controlled by external stimuli, producing more regular and repeatable patterns that can be reliably synchronized during the acquisition cycle while phase maps are acquired at different phases of the motion cycle[7].

When applied to the skeletal muscle, dynamic MRI has proven valuable for quantifying muscle strain and detecting functional differences between populations[8], [9], [10]. Studies have demonstrated its ability to confirm age-related differences in muscle contractile patterns between young and older subjects[11]. Additionally, PC MRI-derived strain measurements have successfully differentiated healthy controls from patients with a range of metabolic neuropathies including facioscapulohumeral muscular dystrophy (FSHD), highlighting its potential as a biomarker for neuromuscular diseases[5], [6].

A notable challenge in PC-MRI is the long acquisition time, which is primarily due to the need for repeated measurements to encode velocity and phase information across multiple directions and time points[12]. Furthermore, the velocity sensitivity needed for muscles is much higher than the one for blood flow leading to longer TE, TRs and stronger gradients, additionally prolonging the measurement time. This is especially problematic in dynamic muscle studies, where high temporal resolution is necessary to accurately capture the motion cycle and total scan time is limited by the onset of muscle fatigue.

To address this challenge, advanced acceleration techniques have been developed. These include methods such as parallel imaging, kt-GRAPPA[13], [14] and compressed sensing[15], [16]. Compressed sensing leverages image sparsity which can be achieved in a variety of schemes and was applied to MSK imaging in both a 3D and 2D acquisitions[17], [18], [19]. While these methods are all described in the literature, wide availability of any implementation is lacking. While some vendors offer their own proprietary accelerated implementations, a true open and documented reference standard is not available.

In this work, we propose a fully open-source pipeline comprising an accelerated 4D flow PC-MRI sequence developed in the vendor agnostic framework PyPulseq[20], [21], data reconstruction with BART[22], and strain data analysis (see Figure 1). The data flow is based on open, vendor-agnostic community standards (ISMRMRD [23] for raw data, and ORMIR-MIDS [24] for image data) and the pipeline preserves geometrical information for interoperability with other anatomical acquisitions. This open-source approach allows implementation across a wide range of MRI systems from multiple vendors, independent of proprietary software costs. We demonstrate this flexibility by presenting forearm and thigh data acquired at two different sites and processed using the same pipeline.

## Methods

# MRI sequence

### Accelerated three-dimensional three-directional phase contrast implementation

A velocity-encoded 4D flow sequence, using 4-point velocity encoding[25] (see Figure 2), was implemented in PyPulseq (v1.4.3) by integrating flow-sensitive gradients along all three physical axes into a conventional cartesian cine triggered gradient echo sequence. These gradients were optimized using the gradient optimization library GrOpt [26] to replace the phase/partition encoding gradients and the readout prephaser by satisfying the zeroth moment (M0) and first moment (M1) constraints to achieve the desired k-space encoding and flow sensitivity while minimizing the echo time (TE).
These waveforms underwent an additional smoothing pass to maintain the slew rate within hardware-safe limits and prevent peripheral nerve stimulation; furthermore, to mitigate the high computational overhead of the 4D acquisition, caching of the gradient objects was utilized during the compilation process.

In order to reduce the total acquisition time the sequence was accelerated with poisson disc undersampling (US, factor 9) along the phase and slice encoding plane, while keeping a fully sampled center of dimensions 13 by 10 points and elliptical scanning by trimming the edges of kspace (see Figure 2).

### Gradient Probing

Different vendors might use different conventions with respect to the physical orientation of gradient polarities, which might result in motion in a physical direction to be encoded either as a positive or negative phase shift according to the internal convention.
To account for these differences, a “gradient probing” sequence was implemented in PyPulseq and based on a dual-echo 3D GRE by adding a small "probing gradient" between the two echoes (either along the readout, phase, or slice directions) to induce a net position-dependent phase advance but minimal intravoxel dephasing (Figure 3). The phase difference between the two echoes was used to evaluate the physical gradient direction, by observing in which direction the induced phase advance was positive or negative. To eliminate phase shifts due to field inhomogeneities, a B0 map with the same echo times as the probing sequence (but without gradient between the two echoes) was acquired and subtracted from the results above. The sequence is meant to be run once per patient orientation to identify the gradients’ direction and correctly interpret the velocity data from the phase contrast imaging sequence.

## Reconstruction

The compressed sensing reconstruction of the undersampled data is achieved using the BART toolbox[22]. ESPIRiT coil sensitivities [27] are estimated and used in the main image reconstruction performed by PICS (Parallel Imaging Compressed Sensing). This reconstruction utilizes L1-regularization applied via a Wavelet sparsity transform in space (x, y, z). After reconstruction, x,y,z velocities are calculated by taking the complex phase difference between the flow-encoded images and the reference image, and converting this difference to physical velocity units using the velocity encoding (VENC) value. The velocity map is then saved in the NIFTI[28] format by using the affine transform extracted from the ISMRMRD [23], to allow registration with anatomical images for Region Of Interest ROI calculation. For the correct interpretation of flow direction and values, this part of the pipeline requires the user to provide a JSON file containing information about the physical direction of the gradients (obtained from the "gradient probing" sequence) and other sequence parameters necessary for the correct calculations of strain values. The parameters defining the size of the fully sampled k-space center and the VENC value are fed to the reconstruction script via a command line interface. A sidecar JSON header is also provided according to the ORMIR-MIDS standard for pipeline-agnostic interpretation of the data.

## Strain analysis

After the velocity calculation, displacement is integrated over time by multiplying the measured velocity field by the $\Delta t$ between the temporal phases. Subsequently, the strain eigenvalue calculation involves computing the deformation gradient tensor from the 3D displacement fields, then deriving the Eulerian strain tensor and extracting its eigenvalues which represent principal strains in three orthogonal directions (stretching, intermediate, and compression components). Subsequently strain rates are extracted by fitting time-varying strain curves with sigmoid functions to quantify how quickly the tissue deforms during the contractions[6].

# Validation Across Two Sites

MRI data acquisition was performed in the forearm muscles of 9 healthy subjects (5 males and 4 females, age 25–45), and in the thigh of 10 healthy male volunteers while the muscles were periodically contracting due to NMES-induced stimulation. All in vivo studies were conducted in compliance with institutional ethics approval and in accordance with the Declaration of Helsinki. Written informed consent was obtained from all participants.The forearm acquisition was performed on a 3T Prisma system (Siemens Healthineers, Erlangen, Germany) at the University Hospital Basel, Switzerland, while the thigh acquisitions were conducted on a 3T Vida Fit system (Siemens Healthineers, Erlangen, Germany) at the Hospital in Koper, Slovenia, allowing validation of the protocol across two MRI platforms.

For the forearm acquisition the protocol included a T1-weighted Turbo Spin Echo (TSE) sequence used for ROI placement, and an additional T2-weighted TSE was acquired before and immediately after stimulation to confirm activation of the targeted muscles. For the thigh acquisition the ROI placement was based on Dixon acquisition. Additional sequence parameters are shown in Table 1.

# NMES Stimulation

The CINE acquisition was triggered with an Neuromuscular Electrical Stimulation (NMES) device (EM 49, BeurerGmbH, Germany), with a train of bipolar rectangular pulses with 300 μs pulse duration at a frequency of 60 Hz, each contraction cycle lasting 1.5 s (0.75 s on and 0.75 s off) [29], [30]. To ensure reproducibility of the NMES stimulation, electrodes (2 5.1 × 8.9 $cm^2$ from TensUnits.com, USA) placement was standardized across all participants. The lower electrode was positioned distally to the distal wrist crease, while the proximal electrode was placed midway along the forearm between the palpable bony landmarks of the processus styloideus ulnae and the medial epicondyle to ensure comparable muscle activation.

For the thigh experiments, the vastus lateralis muscle was targeted. Participants were asked to briefly contract their quadriceps to help visually and palpably identify the belly of the vastus lateralis. The stimulation electrodes were then positioned directly over the activated muscle region to ensure effective recruitment of the muscle during NMES.

## Results

The gradient probing sequence was run for a supine head-first (HFS) sagittal acquisition. The results confirmed that, under this configuration, the slice-encoding gradient increments from left to right, the phase-encoding gradient increments from posterior to anterior, and the readout gradient increments from feet to head (Figure 4). For a feet-first (FFS) configuration, the results differed: left, anterior, and feet directions were all inverted. These directions were subsequently used to correctly assign the sign of the velocity-encoded phase in the 4D flow sequence, ensuring that positive phase shifts correspond to flow in the defined positive physical directions.

The implementation of compressed sensing acceleration into the MR acquisition successfully reduced scan times from 35 and 80 minutes to 5 and 11 minutes for forearm and thigh acquisitions, respectively. This substantial reduction in acquisition time enabled practical in vivo measurements during NMES stimulation, which would otherwise be limited by muscle fatigue and patient tolerance.

Figure 1 shows velocity maps from exemplary acquisitions of the forearm, obtained from the pipeline with application-specific parameters. The muscle regions targeted by NMES (the flexor digitorum superficialis in the forearm) clearly show hyperintensity in the velocity maps, indicating elevated velocities during contraction. A clear hyperintensity is also visible in the t2 weighted TSE after the exercise for the forearm acquisitions (see Figure 5).

The reconstruction and analysis components of the pipeline enabled quantitative assessment of muscle function. Strain maps (Figure 6) derived from the velocity data were used to calculate strain curves across all movement phases; an interactive 3D visualization is available as supplementary material. These curves were fitted with a sigmoid function, allowing for the quantification of contractile parameters including the build-up rate [11] of strain during muscle activation. The first eigenvalues of the strain tensor over the temporal phases are shown in Figure 7 for both applications. Peak and mean strain values were also extracted from this eigenvalue for both applications. Strains measured in the vastus lateralis were approximately one order of magnitude higher than those observed in the flexor digitorum superficialis, with median peak strain of 0.49 vs. 0.063, mean strain of 0.31 vs. 0.031, and buildup rate of 0.030 $s^{-1}$ vs. 0.003 $s^{-1}$ for the leg and arm respectively.

## Discussion and Conclusion

We developed a fully open-source pipeline for quantifying muscle strain using PC MRI during NMES. It works across various anatomical regions and MRI systems, showing that the approach is broadly feasible. By using compressed sensing acceleration, we were able to reduce acquisition times to 5 and 11 minutes for volumetric acquisitions of the forearm and of the thigh, respectively.

In the forearm, the increase in $T_2$ values reflected higher metabolic activity triggered by the electrical stimulation. This change indicates that the muscles were actively responding to the NMES, consistent with greater local energy use [31], [32], [33].

In PC-MRI assessments during NMES, the muscles of the thigh demonstrated higher peak strain magnitude than the muscles of the upper forearm, attributable to its longer fiber length[34] and greater force-generating capacity for knee extension, and partial, contrasting the forearm's finer motor tasks.

Differences in the temporal patterns of strain eigenvalues between the forearm and thigh can be partly explained by the defined limb position during scanning, which affects both baseline muscle tone and contractile behavior[35]. The forearm muscles were scanned in a relaxed state, while the vastus lateralis was slightly contracted because the leg rested on a pillow at a 30° angle. This mild contraction led to a faster strain buildup in the thigh, seen as a steeper rise in the eigenvalue curve.
To facilitate the interpretation of strain eigenvectors, integration of diffusion tensor imaging (DTI) into the current framework would be beneficial [36]. While our current pipeline quantifies strain magnitude along principal directions, DTI could add fiber orientation information, allowing alignment of strain measures with the underlying muscle architecture. This would make the results more physiologically meaningful by distinguishing deformation parallel versus perpendicular to muscle fibers, since muscles have direction-dependent mechanical properties. Accordingly, integrating DTI-derived fiber data into the analysis pipeline is a meaningful next step.

Finally, the proposed pipeline worked smoothly across two different MRI sites, proving the approach to be scanner-independent. This cross-platform compatibility helps overcome limited reproducibility due to proprietary sequences, one of the main obstacles in neuromuscular imaging research.

## Data availability statement

The open-source sequences and analysis pipeline is available at https://github.com/BAMMri/Open4DFlow/tree/v0.1.

## Acknowledgements

SNF grant number 32003B_219674, and Slovenian Research and Innovation Agency, grant numbers: P5-0381 and J5-4593

## Figures

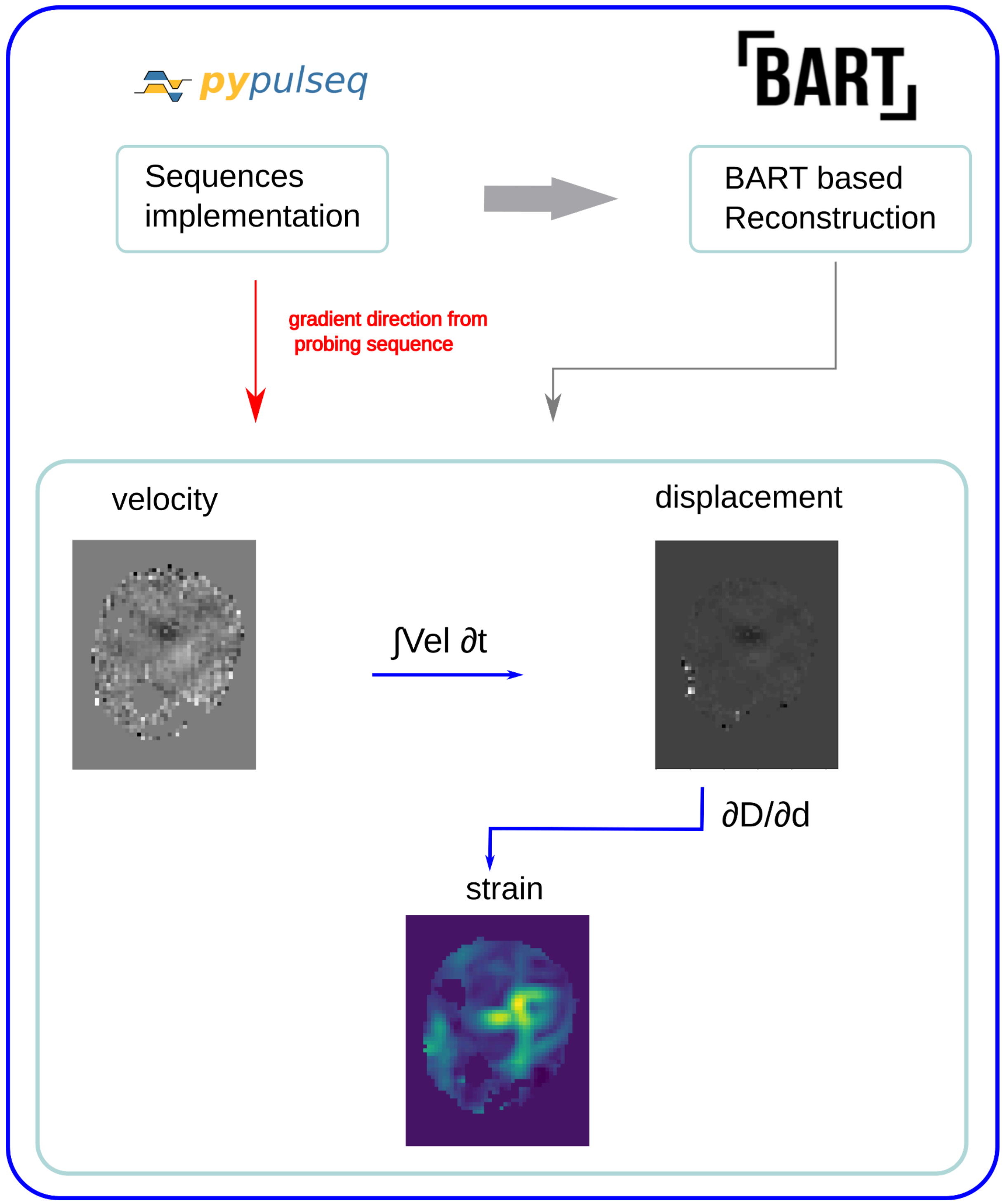

Figure 1: Overview of the data analysis pipeline for 4D Flow and tissue characterization. The workflow starts with the sequence implementation and acquisition, followed by data reconstruction from the sparse k-space samples. The final stage involves the extraction of velocity, displacement, and strain. Representative parametric maps for these parameters are displayed for a healthy 30-year-old female subject.

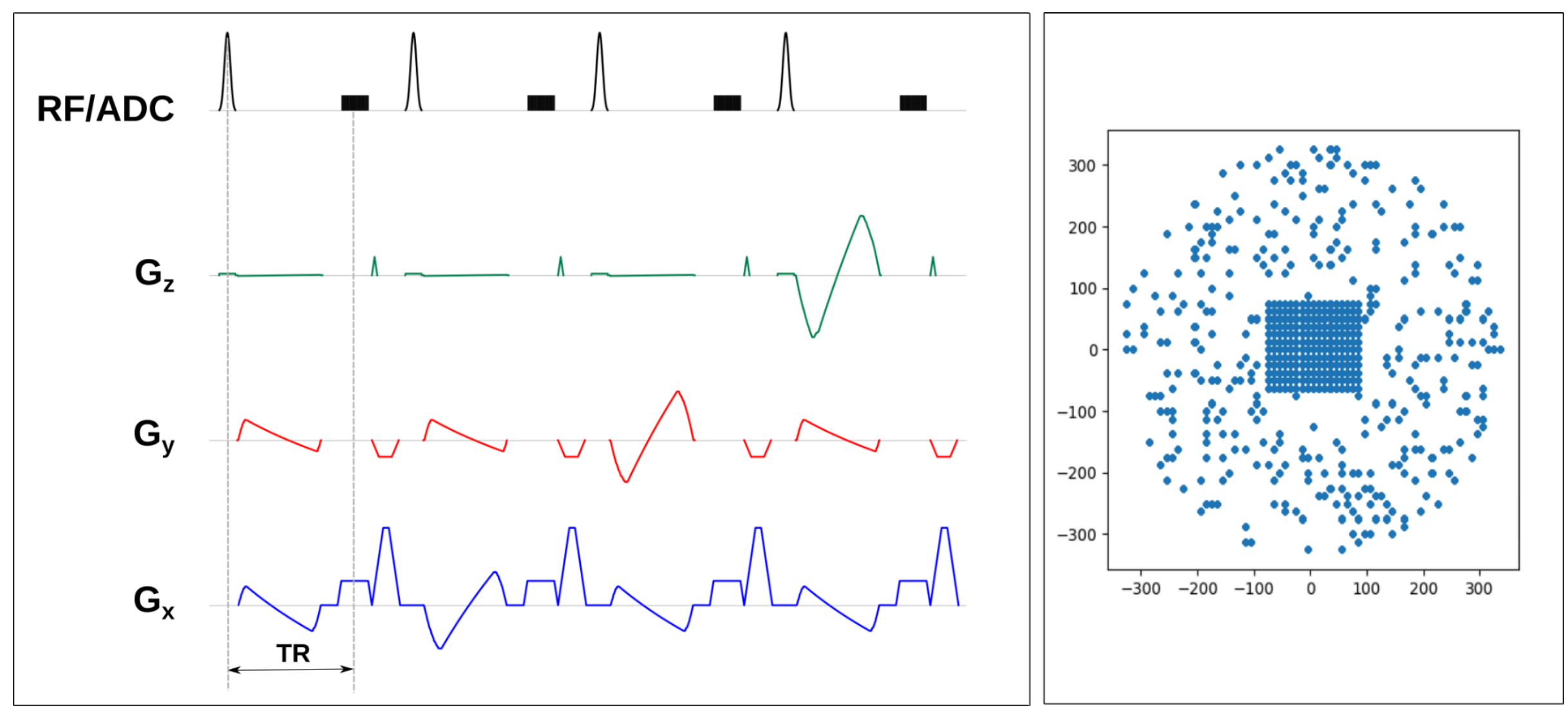

Figure 2: (Left) Pulse sequence diagram of the 4D Flow MRI acquisition. The sequence is based on a 3D Gradient Recalled Echo (GRE) framework, integrated with bipolar flow-encoding gradients along all three spatial axes. These additional gradients enable the quantification of time-resolved, three-dimensional blood flow velocities. (Right) Representative k-space point distributions for imaging of the forearm. Note the implementation of elliptical sampling mask, where the region at the center of k-space is fully sampled, transitioning to a sparse, undersampled pattern toward the edges of k-space.

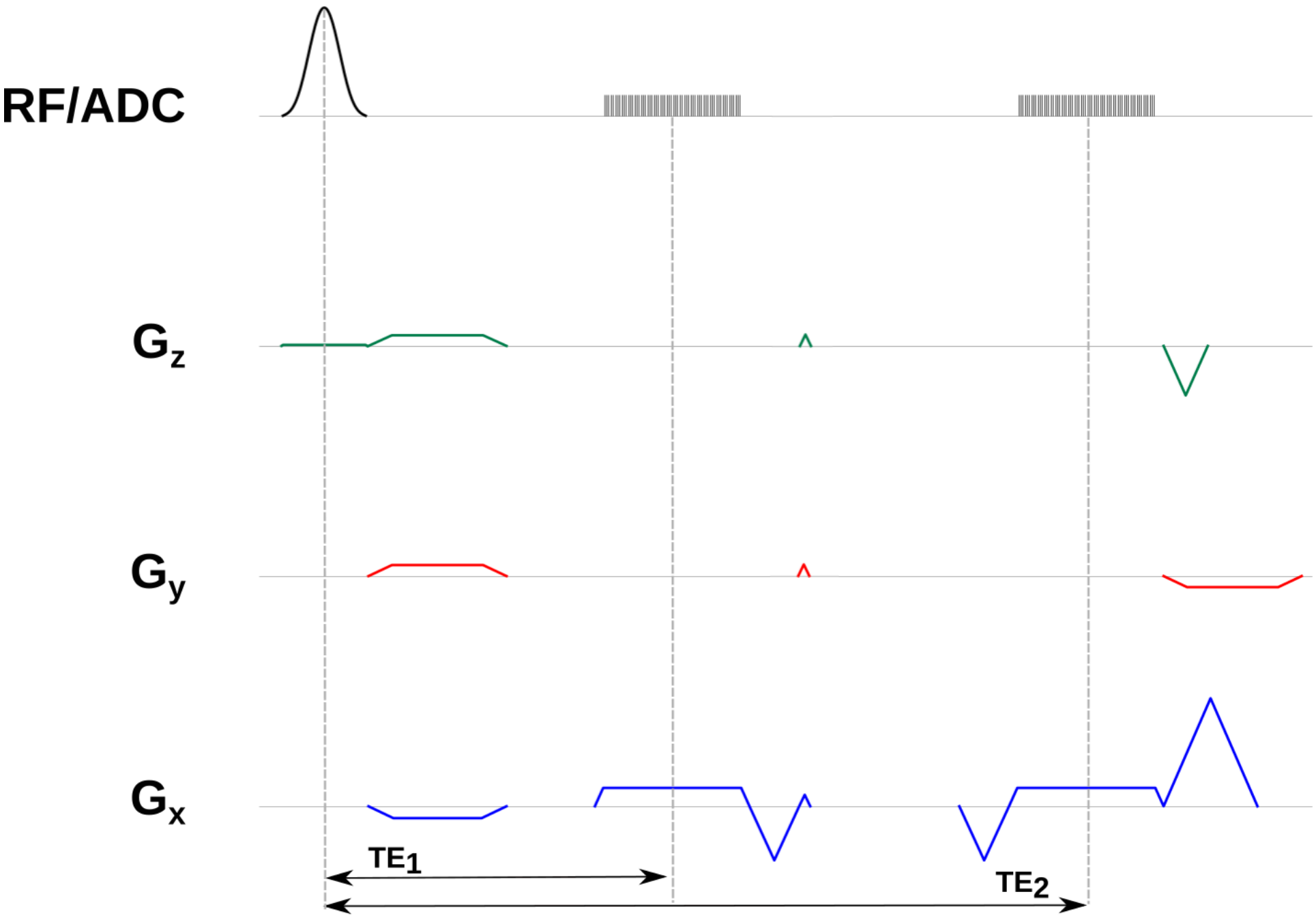

Figure 3: Gradient Probing sequence: dual-echo 3D GRE with an added probing gradient between echoes cycling along the three directions to determine the physical polarity of readout, phase, and slice gradients.

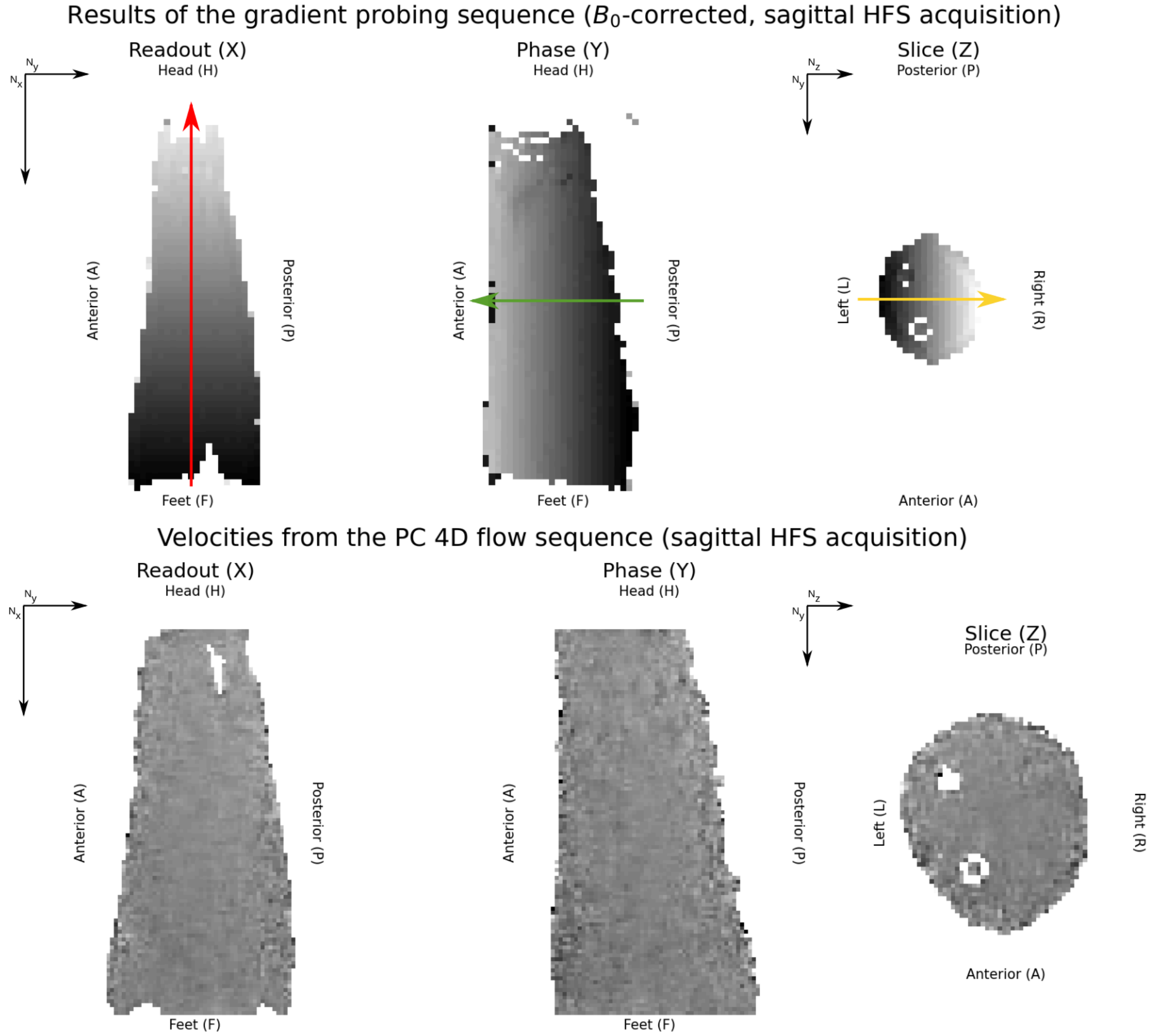

Figure 4: Phase maps obtained from the gradient probing sequence acquired on a forearm phantom in a HFS sagittal orientation, shown for each of the three encoding directions (readout, phase, and slice). The spatial progression of the induced phase advance (from negative to positive) indicates the positive physical direction of each gradient axis under this scanner convention. Arrows denote the identified positive directions.

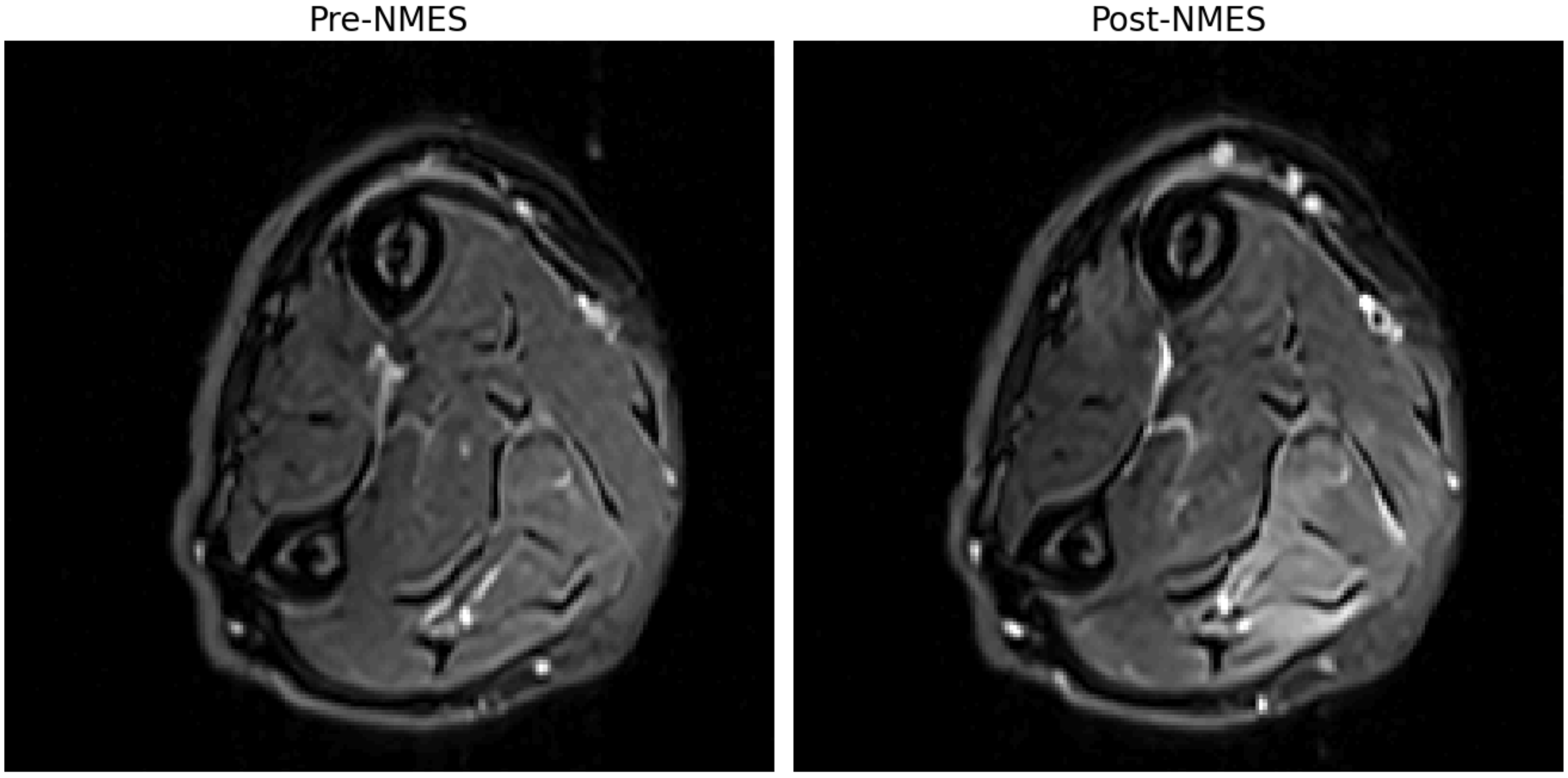

Figure 5: Axial T2-weighted Turbo Spin Echo (TSE) images of a 30-year-old female subject acquired at baseline (left) and post-NMES (right). A distinct region of hyperintensity is visible in the flexor digitorum muscles following exercise. This signal increase is attributed to the increased metabolic activity and extracellular fluid accumulation following neuromuscular electrical stimulation.

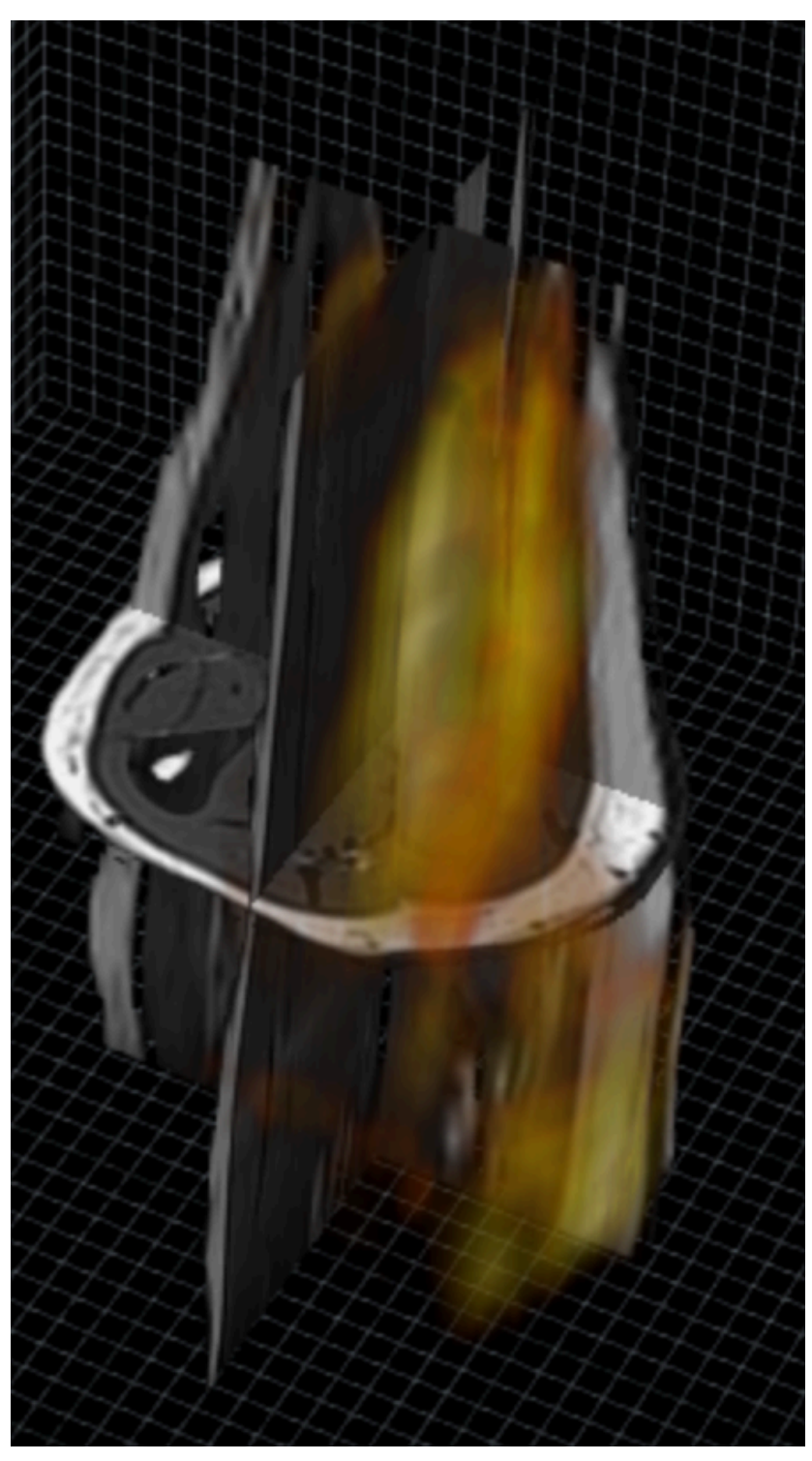

Figure 6: 3D composite rendering of the first eigenvalue ($\lambda 1$) of the strain overlaid on intersecting axial, sagittal, and coronal MRI images from the t1 TSE acquisition for a male volunteer (34 years old). Significant strain is localized along the entire volume of the flexor

digitorum superficialis, with peak strain concentrated in regions targeted by NMES stimulation.

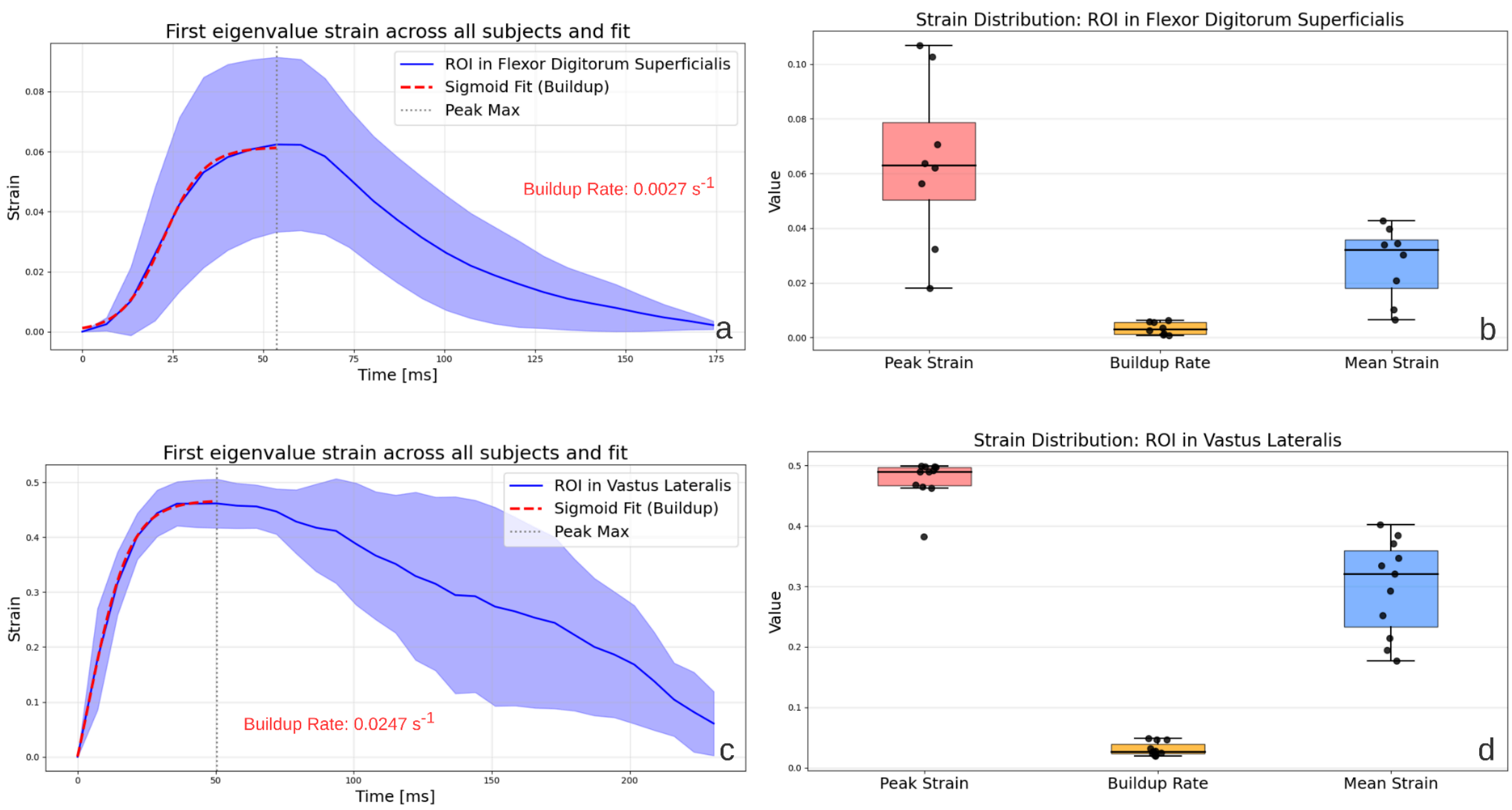

Figure 7: Forearm (a,top left) and Thigh (c, bottom left) first eigenvalue strain (blue) with ±1 SD (shaded) over the temporal phases. A sigmoid fit (red dashed) models the buildup phase to the peak (gray dotted), the mean buildup rate over all the subjects is reported in red. The leg model includes a vertical offset to account for non-relaxed baseline tension, whereas the arm begins at y≈0. top left and bottom right (b,d): Peak and mean strain distribution across muscles and subjects (each dot represents one subject).

## Tables

| Application | Forearm | Leg |
|---|---|---|
| MRI scanner | Prisma (3T) | Vida Fit (3T) |
| TR | 6.7 ms | 7.4 ms |
| TE | 4.5 ms | 4.7 ms |
| FA | 10 deg | 10 deg |
| VENC | 16 cm/s | 25 cm/s |
| Resolution | 1.5x1.5x1.5 $mm^3$ | 2.5x2.5x2.5 $mm^3$ |
| N of Phases | 26 | 33 |
| TA | 5 min | 11 min |

Table 1: Summary table for the different acquisition parameters of the phase contrast sequence for the forearm and leg acquisitions.